\documentclass{jetpl}
\twocolumn

\lat
\usepackage[dvips]{graphicx}
\usepackage{epsfig,color}

\title{Ordering of Fe and Zn ions and magnetic properties of FeZnMo$_3$O$_8$}

\rtitle{Ordering of Fe and Zn ions and magnetic properties of FeZnMo$_3$O$_8$}

\sodtitle{Ordering of Fe and Zn ions and magnetic properties of FeZnMo$_3$O$_8$}

\author{S.\,V.\,Streltsov$^{a,b,}$\/\thanks{e-mail: streltsov@imp.uran.ru},
D.-J.\,Huang$^{c}$
I.\,V.\,Solovyev$^{b,e}$
D.\,I.\,Khomskii$^{d}$}

\rauthor{S.\,V.\,Streltsov, D.-J. Huang, I.\,V.\,Solovyev, D.\,I.\,Khomskii}

\sodauthor{Streltsov, Huang, Solovyev, Khomskii}

\address{$^a$Ural Federal University, Mira Str. 19, 620002 Ekaterinburg, Russia
~\\$^b$Institute of Metal Physics, Russian Academy of Science, S. Kovalevskaya Str. 18, 620041 Ekaterinburg, Russia
~\\$^c$National Synchrotron Radiation Research Center, Hsinchu 30076, Taiwan
~\\$^d$National Institute for Materials Science, MANA, 1-1 Namiki, Tsukuba, Ibaraki 305-0044, Japan
~\\$^e$II. Physikalisches Institut, Universit$\ddot a$t zu K$\ddot o$ln,
Z$\ddot u$lpicher Stra$\ss$e 77, D-50937 K$\ddot o$ln, Germany}

\dates{\today}{*}

\abstract{
In the present paper electronic, magnetic, and structural properties 
of a novel system FeZnMo$_3$O$_8$ with a polar crystal structure are investigated using GGA+U calculations. It is shown that Fe ions preferably occupy octahedral and Zn ions tetrahedral positions. This structural feature is caused by different ionic radii of these ions and not by the exchange coupling. The calculated exchange constants naturally explain magnetic structure observed in this material.
}


\begin{document}

\maketitle

\begin{figure}[b!]
 \centering
 \includegraphics[clip=false,width=0.45\textwidth]{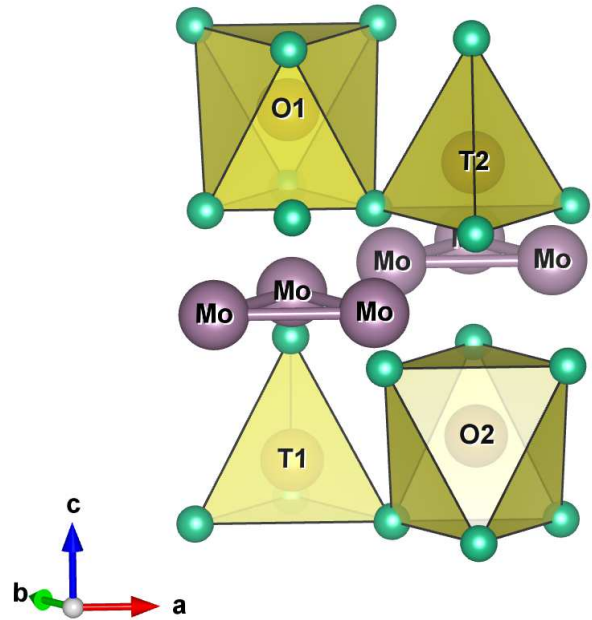}
\caption{\label{crystal-structure}Fig. 1. Crystal structure of FeZnMo$_3$O$_8$. Mo and O ions are shown in violet and light green respectively. There are two possible octahedral (O1 and O2) and two tetrahedral (T1 and T2) positions of Fe and Zn ions.}
\end{figure}

Coupling of electricity and magnetism in magnetoelectric and multiferroic materials present significant interest both from the fundamental physical point of view and for practical applications, e.g., in spintronic devices. Therefore a search and study of novel materials with interesting magnetoelectric properties is very active nowadays. A novel interesting class of these systems is given by the materials of the type M$_2$Mo$_3$O$_8$ (M=Mn, Fe, Co, Ni, and also Zn). They have polar crystal structure (pyroelectric group), shown in Fig.~\ref{crystal-structure}, with two types of positions for metals M, octahedral (O1 and O2) and tetrahedral (T1 and T2).  Magnetoelectric effect in these materials is very interesting: it strongly depends on composition, e.g., it changes sign at substitution of Mn in Mn$_2$Mo$_3$O$_8$  by Fe \cite{Kurumaji-2017b}, and can be strongly enhanced by magnetic field \cite{Wang-2015}. Magnetic structure of different materials of this class is also quite flexible: thus the substitution of Fe in Fe$_2$Mo$_3$O$_8$  by Zn leads to the change of magnetic ordering from antiferro- to ferromagnetic \cite{Varret-1972,Kurumaji-2017a}. The nature of this behavior is still not clarified. To unravel it, we carried out extensive theoretical study, using ab-initio methods, combining those with the spectroscopic study (preliminary results of these experiments were presented at the APS March meeting in 2019\cite{HuangAPS}). 


One of the main problems in mixed materials of this class, e.g. in (Fe$_{1-x}$Zn$_x$)$_2$Mo$_3$O$_8$ is the distribution of constituent ions among inequivalent sites, octahedral and tetrahedral ones~\cite{Varret-1972,Czeskleba-1972}. Also the magnitude and the sign of exchange interactions in these materials is not clear a priori. As one sees from the crystal structure, shown in Fig. 1, there are several inequivalent exchange passes even for nearest neighbour interaction: in-plane $J_{OT}^{ab}$ and interlayer $J_{OO}^c$, $J_{TT}^c$ and $J_{OT}^c$ exchanges. Moreover, the ``vertical'' exchanges $J_{OT}^c$ might be also different, because of the orientation of MO$_4$ tetrahedra (which all point in the same direction, this is in fact what makes these systems polar). 
Since Fe$_2$Mo$_3$O$_8$ is known to be a multiferroic, i.e. a material in which electronic, magnetic, and structural properties are strongly coupled, one might expect that exchange interaction may affect crystal structure of doped (Fe$_{1-x}$Zn$_x$)$_2$Mo$_3$O$_8$ as well. 

The bulk experimental data show that the substitution of Fe by Zn in Fe$_2$Mo$_3$O$_8$ changes magnetic ordering from antiferromagnetic to ferromagnetic, in particular for half-doping $x=1$, in FeZnMo$_3$O$_8$\cite{Varret-1972,Kurumaji-2017a}. Our preliminary spectroscopic data \cite{HuangAPS} point toward preferential location of Zn in tetrahedral sites; this also agrees with the old estimates \cite{Bertrand-1975}. If true, this would make the magnetic subsystem in FeZnMo$_3$O$_8$ relatively simple: only octahedral sites would be occupied by magnetic ions Fe$^{2+}$, so that only one type of exchange, the diagonal interlayer Fe-Fe exchange $J_{OO}^c$ remains, and if this would be ferromagnetic, it could explain the observed magnetic behavior. However, from the existing experimental data one cannot yet make definite conclusions in this respect, although they indeed point in this direction.

We used pseudopotential VASP code to calculate electronic, magnetic, and structural properties of FeZnMo$_3$O$_8$ \cite{vasp}. The Perdew-Burke-Ernzerhof version of exchange correlation potential was utilized \cite{Perdew1996}.  Following radii of atomic spheres were chosen: $R_{Fe}=1.302$\AA, $R_{Mo}=1.455$\AA, $R_{O}=0.820$\AA,  and $R_{Zn}=1.270$\AA. We used GGA+U approach to take into account strong electronic correlations~\cite{Liechtenstein1995}. On-site Hubbard $U$ and intra-atomic exchange $J_H$ for Fe ions were chosen to be 6.0 and 1.0 eV\cite{Streltsov2017}. $5 \times 5 \times 3$ mesh in the $k$-space was used in all calculations. The crystal structure was taken from Ref.~\cite{Wang-2015}. Ionic relaxations were performed until total energy difference between two ionic iterations 
was larger than 10$^{-5}$ eV. Integration of the electronic spectrum was performed using  Gaussian smearing (with smearing parameter 0.1 eV). Exchange parameters for Heisenberg model were calculated using JaSS code\cite{jass}.
\begin{figure}[t!]
 \centering
 \includegraphics[clip=false,width=0.49\textwidth]{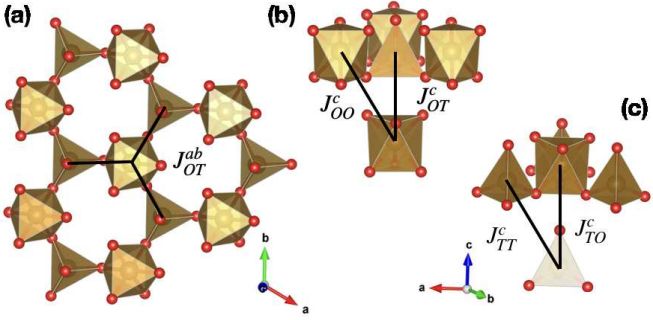}
\caption{\label{exchanges}Fig. 2. Different possible exchange coupling parameters in M$_2$Mo$_3$O$_8$ type of structures.}
\end{figure}

We start our investigation by studying effect of exchange interaction on distribution of Fe and Zn ions among octahedral and tetrahedral positions. In particular we calculated exchange parameters in FeZnMo$_3$O$_8$ for various occupations of four available positions in the unit cell by two Fe ions keeping the crystal structure untouched. Finding total energy difference between ferromagnetic (FM) and antiferromagnetic (AFM) configurations we obtain that $J_{OT}^{ab}=0.96$ meV, $J_{OT}^{c}=-0.61$ meV, $J_{TT}^{c}=0.03$ meV, and $J_{OO}^{c}=0.01$ meV, if Heisenberg model is written in the form 
\begin{eqnarray}
H = \sum_{i \ne j} J_{ij} \mathbf S_i \mathbf S_j. 
\end{eqnarray}
Thus, one may see from these results that exchange interaction would stabilize configuration with Fe ions occupying equally tetrahedral and octahedral positions, but concentrated in the same plane (to maximize gain in magnetic energy due to $J_{OT}^{ab}$). 


Next we relaxed the crystal structure of FeZnMo$_3$O$_8$ for different Fe and Mo distributions in two possible magnetic orders (FM and AFM).  Results are summarized in Tab.~1. First, one may see that the state with all Fe ions occupying octahedral positions has the lowest total energy. Moreover, the energy of any 50:50 configuration (50\% Fe is octahedral and 50\% Zn is in tetrahedral positions) is  equal to about half of the energy difference between configurations, where all Fe ions are in octahedral positions. This means that this is not magnetic energy, but lattice distortions (i.e. elastic energy), which decides which positions are occupied by Fe ions. Quite naturally, the energy scales of magnetic and elastic interactions differ by about three orders of magnitude.

It is interesting that the total energy calculations also reproduce the ground state magnetic structure: experimentally FeZnMo$_3$O$_8$ is indeed ferromagnetic. However, strictly speaking one cannot extract exchange constants from the total energies presented in Tab. 1, since the crystal structure was relaxed independently for each type of magnetic order (and for each type of ionic distribution as well). In order to calculate exchange parameters in FeZnMo$_3$O$_8$ we relaxed both lattice parameters and ionic positions in non-magnetic GGA calculations and used this (equally good for FM and AFM orders) structure to obtain $J_{OO}^c = -0.2$ K. Thus with this distribution of ions (Fe in octahedra, Zn in tetrahedra), with  the only remaining exchange interaction $J^c_{OO} = -0.2$ K,  we naturally obtain the ferromagnetic ordering in FeZnMo$_3$O$_8$, observed experimentally\cite{Varret-1972}.
\begin{table}[t]
\caption{Tab. 1. Energies (per formula unit) of different distributions of Fe and Zn ions among octahedral (O1 and O2) and tetrahedral (T1 and T2) positions, see Fig.~\ref{crystal-structure}, as obtained after optimization of the ionic positions in the GGA+U calculations.\label{Energy-table}}
 \begin{tabular}{lccccc}
  \hline \hline
O1  & O2 & T1  & T2 & Magnetic order & Energy \\
 \hline \hline
 Fe & Fe  & Zn & Zn &   FM          & 0 \\
 Fe & Fe & Zn & Zn &  AFM         & 0.0002 eV \\
\hline
Zn & Zn & Fe & Fe  &    FM          & 1.2702 eV \\
Zn & Zn & Fe & Fe  &   AFM         & 1.2698 eV \\ 
\hline
Fe & Zn & Fe & Zn  &    FM          & 0.6405 eV \\
Fe & Zn & Fe & Zn  &   AFM         & 0.6403 eV \\ 
\hline
Fe & Zn & Zn & Fe  &    FM          & 0.6440 eV \\
Fe & Zn & Zn & Fe  &   AFM         & 0.6301 eV \\ 
  \hline \hline
 \end{tabular} 
\end{table} 

Preferable occupations of octahedral positions by Fe and not by Zn ions is due to its larger ionic radius ($R_{Fe^{2+}}^{octa} = 0.78$ \AA, $R_{Zn^{2+}}^{octa} = 0.74$ \AA~  and $R_{Fe^{2+}}^{tetra} = 0.63$ \AA, $R_{Zn^{2+}}^{tetra} = 0.60$ \AA \cite{Shannon-1976}), so that it is natural for larger ions to go in a larger polyhedron. Indeed, if we put smaller Zn to an octahedron it gets more distorted. E.g., the bond angle variance defined as $\langle \sigma^2 \rangle = \sum_{i=1}^{m} (\phi_i - \phi_0)^2/(m-1)$ (here $m$ is the number of bond angles and $\phi_0$ is the bond angle for regular octahedron) increases from 69.1$^{\circ}$, when Fe is in octahedral positions, to 74.4$^{\circ}$, when Zn sits in octahedra.

\begin{figure}[t!]
 \centering
 \includegraphics[clip=false,width=0.45\textwidth]{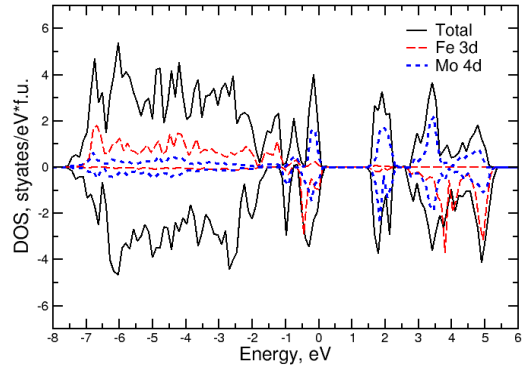}
\caption{\label{DOS}Fig. 3. Partial density of states (DOS) plot as obtained in the GGA+U calculations for FeZnMo$_3$O$_8$ with Fe occupying octahedral sites. Ferromagnetic order was assumed. The Fermi energy is at zero.}
\end{figure}

For completeness in Fig.~\ref{DOS} we also present the density of state plot for FeZnMo$_3$O$_8$. One may see that both top of the valence and bottom of the conduction bands are formed by Mo $4d$ states. Mo is 4+ in these compounds, i.e. there are two electrons per each molybdenum. Mo ions form trimers - triangular diamagnetic ``molecules'' in M$_2$Mo$_3$O$_8$ structure. This makes treatment of these states particularly complicated. On one hand, these are $d$ states and they are correlated and therefore $U$ correction should be applied for them. On the other hand they form molecular orbitals, but methods such as GGA+U  are essentially one-electron methods (one Slater determinant) and thus cannot correctly describe molecular orbital states, so that
one needs to use more sophisticated methods such as e.g. cluster dynamical mean field theory in this situation (C-DMFT)\cite{Biroli2002,Streltsov2016}. Nevertheless, we checked that an account of Hubbard correlations on Mo ($U=3$ eV, $J_H=0.7$ eV) on the GGA+U level does not change main results of the present work.

To summarise, we studied theoretically the very interesting system  of the novel class M$_2$Mo$_3$O$_8$ (M=Mn, Fe, Co, Ni, Zn) with polar crystal structure, which show diverse magnetic properties and very interesting magnetoelectric effects. Specifically we concentrated on the  mixed materials of this class (Fe$_{1-x}$Zn$_x$)$_2$Mo$_3$O$_8$, for which most experimental studies are done at present\cite{Kurumaji-2017a,Wang-2015,Yu2018,Kurumaji-2017b}. The most important question here is the distribution of Fe and Zn at different positions (octahedral, tetrahedral) existing in these systems. This is very important for all the properties of corresponding systems. We demonstrated that there is very strong preference of Zn to occupy the tetrahedral sites, so that  for 50\%  substitution, in the title compound FeZnMo$_3$O$_8$, all tetrahedral sites are occupied by nonmagnetic Zn, and magnetic Fe ions are all in octahedral positions. This, together with the ferromagnetic octahedral-octahedral exchange, which we calculated theoretically, naturally explains ferromagnetic ordering observed in FeZnMo$_3$O$_8$ experimentally. These results will be also very useful for interpreting spectroscopic data on FeZnMo$_3$O$_8$\cite{HuangAPS}.  Apparently the tendency of relative distribution of different ions in this class of materials, determined, as we demonstrated above, mainly by the size of respective ions, might also work  in other systems of this interesting class.

We thank Alexey Ushakov who did initial calculations for parent material (Fe$_2$Mo$_3$O$_8$). Calculation of the exchange constants were carried by S.V.S. and supported by Russian Science Foundation via project 17-12-01207. D.I. Khomskii was supported by the Deutche Forschungsgemeinschaft (DFG, German Research Foundation), project number 277146847 -- CRC 1238.

\end{document}